\documentclass[twocolumn,showpacs,%
  nofootinbib,aps,superscriptaddress,%
  eqsecnum,prd,notitlepage,showkeys,10pt]{revtex4-1}

\usepackage{amssymb}
\usepackage{amsmath}
\usepackage{graphicx}
\usepackage{dcolumn}
\usepackage{hyperref}
\usepackage{bm}
\usepackage{bbm}
\usepackage{float}
\usepackage{physics}
\usepackage{comment}

\begin{document}

\title{Optimizing Airborne Wind Energy with Reinforcement Learning}
\author{N. Orzan}
\affiliation{The Abdus Salam International Center for Theoretical Physics ICTP, Trieste 34151, Italy}
\affiliation{University of Groningen, 9700 AB Groningen, the Netherlands}
\author{C. Leone}
\affiliation{The Abdus Salam International Center for Theoretical Physics ICTP, Trieste 34151, Italy}
\affiliation{SISSA International School for Advanced Studies, Trieste I-34136, Italy}
\author{A. Mazzolini}
\affiliation{The Abdus Salam International Center for Theoretical Physics ICTP, Trieste 34151, Italy}
\affiliation{Laboratoire de physique de l’École normale supérieure (PSL University), CNRS, Sorbonne Université, and Université de Paris, 75005 Paris, France}
\author{J. Oyero}
\affiliation{The Abdus Salam International Center for Theoretical Physics ICTP, Trieste 34151, Italy}
\affiliation{Nonlinear Physical Chemistry Unit
Service de Chimie Physique et Biologie Théorique
Université Libre de Bruxelles
CP 231 - Campus Plaine, 1050 Brussels, Belgium}
\author{A. Celani}
\affiliation{The Abdus Salam International Center for Theoretical Physics ICTP, Trieste 34151, Italy}

\thanks{Blah}

\begin{abstract}
Airborne Wind Energy is a lightweight technology that allows power extraction from the wind using airborne devices such as kites and gliders, where the airfoil orientation can be dynamically controlled in order to maximize performance. 
The dynamical complexity of turbulent aerodynamics makes this optimization problem unapproachable by conventional methods such as classical control theory, which rely on accurate and tractable analytical models of the dynamical system at hand.
Here we propose to attack this problem through Reinforcement Learning, a technique that -- by repeated trial-and-error interactions with the environment -- learns to associate observations with profitable actions without requiring prior knowledge of the system. We show that in a simulated environment Reinforcement Learning finds an efficient way to control a kite so that it can tow a vehicle for long distances. The algorithm we use is based on a small set of intuitive observations and its physically transparent interpretation allows to describe the approximately optimal strategy as a simple list of manoeuvring instructions. 
\end{abstract}

\maketitle

\section{Introduction}

Airborne wind energy (AWE) is a technology aiming to obtain usable power by means of flying devices \cite{awebook} which has the potential of replacing the traditional towered wind turbine architecture \cite{BECHTLE20191103}. Airborne wind energy systems usually consist in an airfoil providing traction power - generally a kite or a glider -  which is either connected by a tether to a ground station that converts power into electricity by a turbine, or used to tow a vehicle~\cite{CHERUBINI20151461,wellicome}.

There are many advantages of AWE over standard wind turbines: reduced costs for construction; lower environmental impact; saving in building materials; ease in displacing the device \cite{canale07,EU}. Here we focus on one distinguishing feature of AWE that is the possibility of adapting to the rapid and local changes in wind conditions by controlling the orientation of the kite. This allows to maintain the kite airborne and effective under varying wind and weather conditions \cite{ilzhofer}. 

The task of finding the best way of manoeuvring the kite in order to maximise power production has been previously addressed by means of optimal control theory \cite{kitegencontrol,skysailscontrol,williams}, using methods from  Non-linear Model Predictive Control \cite{NMPC}. These are essentially planning algorithms that crucially rely on a dynamical model in order to predict the future evolution of the system. For AWE this requirement translates in finding an accurate model of both the dynamics of the kite and of the turbulent wind. Such model would involve a huge number of dynamical degrees of freedom and the resulting optimization problem would then become computationally intractable. Moreover, even assuming that the previous difficulties could be overcome, turbulence is famously characterized by its unpredictability, undercutting the power of predictive control. To bypass these difficulties, the effect of wind turbulence is often ignored or too crudely approximated, for instance by adding small uncorrelated fluctuations \cite{ARGATOV,canale07,houska}. The control strategies so obtained may however turn out to be severely suboptimal when put to the test in a realistic turbulent environment.

Here, we propose to use Reinforcement Learning (RL) to find effective control strategies for AWE in a realistic turbulent environment.
The core idea of RL consists in learning how to control a system in order to achieve a long-term goal without relying on detailed \textit {a priori} knowledge of its dynamics and therefore avoiding the shortcomings of predictive control~\cite{sutton}. By repeatedly interacting with the environment that surrounds it, the controller learns by trial and error to associate valuable actions to specific contexts~\cite{littman2015reinforcement}. RL has been shown to be competitive with Model Predictive Control even in situations where an accurate model of the system is available \cite{Ernst}. The ability of RL in finding effective control strategies in dynamic and unpredictable environment such as the turbulent atmosphere has been recently showcased in complex tasks such as thermal soaring and balloon navigation~\cite{reddy2018glider,Bellemare}.

In this article we provide a proof of concept that RL can be successfully used for AWE. We consider a simulated environment that simulates the dynamics of a ship towed by a kite that flies in the turbulent atmospheric boundary layer. A major difficulty of this task is to find appropriate controllers that sense and exploit the variable wind strength and at the same time avert disastrous crashes. We find that a compact RL algorithm based on a small set of physically intuitive observations and controls is able to find effective strategies to tow the vehicle for long distances, efficiently converting the energy of the turbulent wind into directed motion.

\begin{figure*}[ht]
\centering
\includegraphics[scale=0.58]{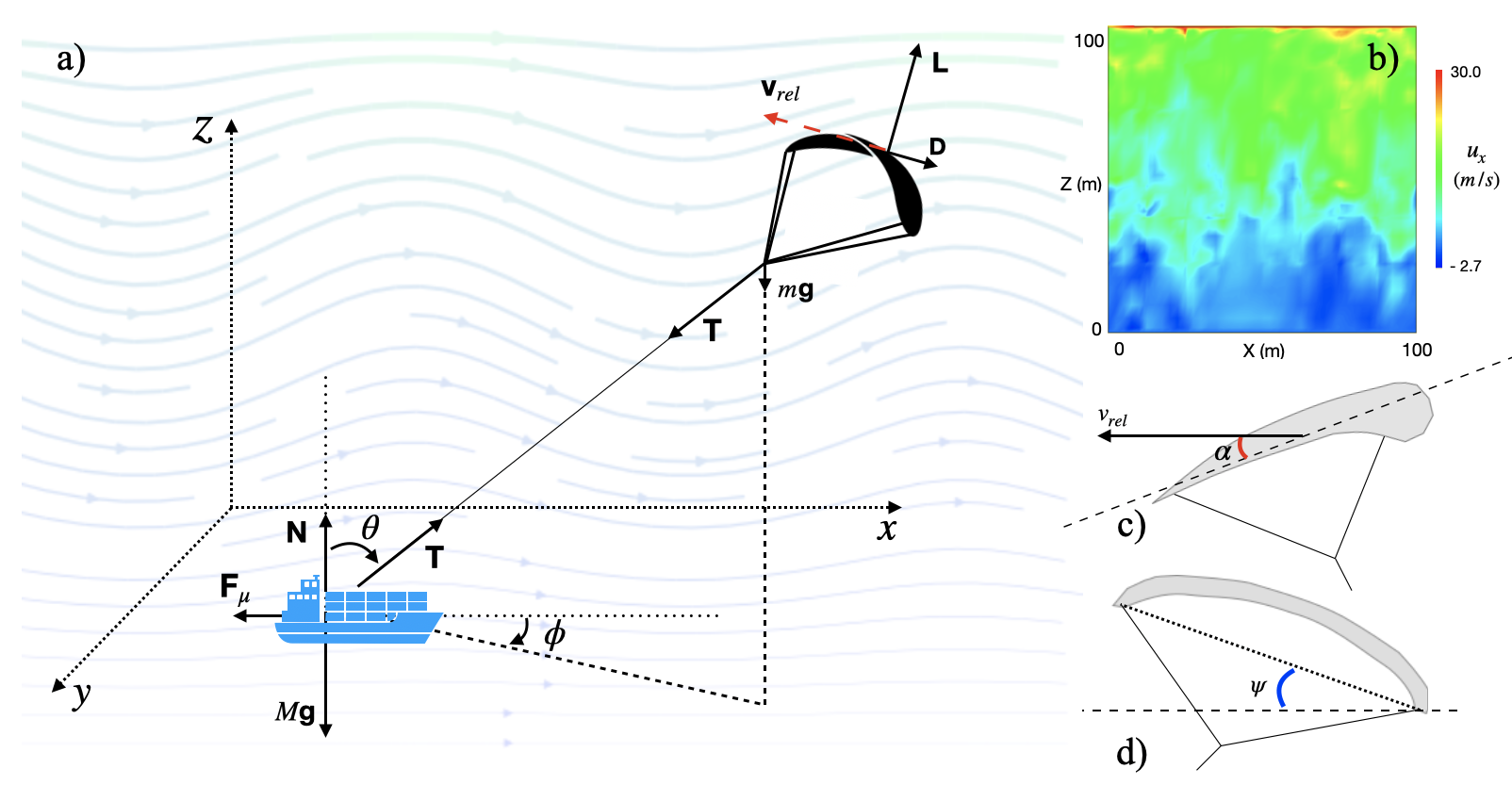}
\caption{ The simulated environment for AWE. a) Sketch of the kite-ship system. b) Snapshot of a vertical cross-section of the horizontal wind velocity in the turbulent flow. c) The attack angle $\alpha$ is the angle between the longitudinal axis of the kite and the relative velocity; its control allows the kite to dive and rise. d) The bank angle $\psi$ changes the direction of the lift force and its control makes the airfoil turn left and right.}
\label{fig:3dimage}
\end{figure*}

\section{The simulated environment}
In this section we describe the simulated environment for AWE production.

\subsection{Kite and vehicle dynamics} 
We consider the system composed by a kite connected to a vehicle (e.g. a ship) through an inextensible cable.  The vehicle can move on the surface, while the kite is able to travel on the whole space above it. Sagging of the rope is neglected as the cable is understood to be always in tension.

The kite has mass $m$ and is subject to the action of the wind, summarized by the total aerodynamic force $\bm{F}^{aer}$, the gravitational force $m \bm g$ and the tether tension $\bm T$. The vehicle of mass $M$ is subject to the gravitational force $M \bm g$, the same tension $-\bm T$ acting on the other end of the rope, the sliding friction force $\bm{F}^{\mu}$ and the normal reaction of the ground/sea surface $\bm N$ (see Fig.~\ref{fig:3dimage}a)

The positions of the kite $\bm{x}_k$ and of the vehicle $\bm{x}_v$ obey the Newton's law
\begin{equation}
    \begin{cases}
    m \bm{\ddot x}_k= \bm F^{aer}+m\bm g-\bm T\\
    M \bm{\ddot x}_v=\bm F^{\mu}+\bm N+M\bm g+\bm T
    \end{cases}
\label{eq:system}
\end{equation}
subject to the constraints of fixed tether length \(\abs{\bm{x}_k-\bm{x}_v} = R\) and that the vehicle remains anchored to the ground \(z_v= 0\).

The friction force has amplitude $F^{\mu} = \mu N = \mu (M g-T_z)$ where $T_z$ is the vertical component of the tension applied to the vehicle. Its direction is opposed to the vehicle velocity $\dot{\bm x}_v$.

The aerodynamic forces are customarily decomposed in \textit{lift} and \textit{drag}: \(\bm F^{aer} = \bm L \,+\, \bm D\). Both components depend on the relative velocity of the kite with respect to the wind \(\bm{v}_{rel} = \bm{v}_k - \bm{u}\), where \(\bm{u}\) is the wind velocity and $\bm{v}_k=\dot{\bm{x}}_k$. 

The amplitudes of lift and drag 
are     $L=\frac{1}{2} \rho A C_L(\alpha) \abs{\bm{v}_{rel}}^2$ and $D=\frac{1}{2} \rho A C_D(\alpha) \abs{\bm{v}_{rel}}^2$ respectively,
where \(\rho\) is the air density, \(A\) is the kite surface area and \(C_L(\alpha)\), \(C_D(\alpha)\) are the lift and drag coefficients. The latter depend on the \textit{attack angle} of the kite \(\alpha\) (fig. \ref{fig:3dimage}c), which is the angle between the longitudinal axis of the kite and the relative velocity. By changing the attack angle the kite can rise or dive. The values of the coefficients are measured empirically for different airfoils and here we used the ones given in Ref.~\cite{canale} for our simulation. 

As for the direction, the drag is anti-parallel to the relative velocity whereas the lift lies in the plane perpendicular to it. Its exact orientation depends on the \textit{bank} angle \(\psi\) (fig. \ref{fig:3dimage}d), which can be controlled and allows for the kite to make turns. Namely, if we take $\bm e_r=({\bm x}_k-{\bm x_v})/R$ to be the unit vector which identifies the direction of the tether and define the two unit vectors perpendicular to the relative velocity
$\bm{e}_t = \bm{e}_{r} \times \bm{v}_{rel}/\abs{\bm{e}_{r} \times \bm{v}_{rel} }$ and
$\bm{e}_n = \bm{v}_{rel} \times \bm{e}_t / \abs{\bm{v}_{rel} \times \bm{e}_t}$
we have
$\bm{L} = L\,(\bm{e}_t \sin(\psi) + \bm{e}_n \cos(\psi))$ (see Ref.~\cite{williams} and the Supplementary Information for more details on the dynamics).

We assume that the kite can be controlled by changing its attack and bank angles - and consequently its orientation in space - by operating on the lines that are connected to the sides of the airfoil.

\subsection{The turbulent wind}
We simulate the turbulent wind flow in the atmospheric boundary layer by solving the incompressible Navier-Stokes equations
\begin{equation}
\begin{cases}
      \frac{\partial \bm{u}}{\partial t}+ (\bm{u} \cdot \nabla) \bm{u}= -\nabla p + \nu \nabla^2 \bm{u}  \\
    \nabla \cdot \mathbf{u}=0  
\end{cases}
    \label{eq:Navier}
\end{equation}
in a cubic domain of side $\approx 100$ m. No-slip boundary conditions are imposed at the bottom and a fixed velocity $U_W=30$ m/s, directed along the streamwise direction $x$, is imposed at the top (fig. \ref{fig:3dimage}b). Periodic boundary conditions are applied in the stream-wise \(x\) and span-wise \(y\) directions. This configuration is usually dubbed a turbulent Couette flow. The Reynolds number \(Re = \frac{U_{W} \delta}{2 \nu}\), where \(\delta\) is the half-height of the channel and \(\nu\) is the kinematic viscosity, is \(Re = 65610\) which indicates the presence of a turbulent flow. The simulation uses a Spectral Element Method as implemented in the free CFD software Nek5000 \cite{patera}.
At the steady state, the statistical properties of the flow are consistent with those observed in previous works \cite{avsarkisov} (see the Supplementary Information for more details).

\section{Learning to control the kite}
As anticipated in the Introduction, we have approached the problem of maximising the traction power of a kite in a turbulent flow by means of Reinforcement Learning. In this Section we give a brief description of the algorithm that we have used and how it has been implemented in the case at hand.

\subsection{Reinforcement Learning}
The objective of RL is to optimize the performance of a controller, or agent, in a generic goal-directed task (see Ref.~\cite{sutton} for a comprehensive introduction to the subject).
At each time step, the controller receives an observation $S$ that provides some information about the instantaneous, usually hidden, full Markov state of the system $X$ and takes an action \(A\) according to a \textit{policy} \(\pi\), that is a probability distribution over actions which depends on the history of previous observations. The system then evolves in a new state $X'$ and the process repeats itself until the end of an \textit{episode} of duration $T$. At each time step, the agent receives a reward, or reinforcement signal, $R$.
The objective is to find a policy that maximizes the expected sum of future rewards from a given time $t$, known as the \textit{return} \(G_t = \sum_{k=t}^T R_k\).

Importantly, the agent does not need any \textit{a priori} knowledge of the dynamics of the hidden Markov states $X \to X'$ and of the structure of the rewards. It has to learn how to associate past observations with profitable actions just by interacting with the environment.

Finding a policy that depends on the full history of observations is however computationally very demanding. Here we will settle for the comparatively less ambitious goal of searching for good \textit{reactive policies}, that is strategies that choose to take an action $A$ based only on the last observation $S$ according to a policy $\pi(A|S)$. If the observations were the actual Markov states for the system, i.e. $S=X$, then the knowledge of past observations would be superfluous and one can consider only reactive policies. However, this nearly never happens in practice.

Approximately optimal reactive policies can be found with limited computational effort
when observations and actions can be represented by a finite discrete set. In this case tabular time-difference learning algorithm such as SARSA (the acronym of State-Action-Reward-State-Action) can find good strategies by interacting with the environment \cite{sutton}.

SARSA starts off with an estimate of the expected return that a policy can achieve in the future if it has received the observation $S$ and executed the action $A$: \(\hat{Q}(S,A)\). Then, at each step the algorithm makes decision according to the current estimate and improves it as follows:
\begin{itemize} 
\item[\textit{i)}] Derive a policy from the current estimate \(\hat{Q}\) with $\epsilon-$\textit{greedy} exploration: given an observation $S$ pick the action which has the largest estimated return $A=\mathrm{argmax}_a \hat{Q}(S,a)$ with probability $1-\epsilon$ and any possible action at random otherwise;
\item[\textit{ii)}] After having observed $S'$ and taken action $A'$ according to the same policy, update the estimate as
\(\hat{Q}(S,A)\leftarrow \hat{Q}(S,A) + \eta  (R+\hat{Q}(S',A')-\hat{Q}(S,A))\). All the other entries of the matrix \(\hat{Q}\) are left unchanged.
\end{itemize}
This procedure is iterated until convergence is obtained.

In the expressions above, the learning rate $\eta$ and the exploration probability $\epsilon$ depend on the number of visits of the current  observed state-action pair. With a proper decreasing schedule of these parameters, SARSA converges with probability one to the optimal control strategy if observations are Markov states for the system. When there is only partial observability, it converges to the best possible reactive strategy for the given choice of observable states.

In order to couch our AWE optimization problem in the language of reinforcement learning we now turn to define the observable states (for short, observables) $S$, the actions $A$ and rewards $R$ for our AWE system. 

\subsection{Actions}
As discussed earlier, the kite can rise or dive by changing the angle of attack $\alpha$ and it can turn by modifying the bank angle $\psi$. Therefore an intuitive and minimal way to control the kite is to decrease, leave unchanged or increase both $\alpha$ and $\psi$ by some fixed amount:
\begin{equation}
\begin{array}{lll}
\alpha \rightarrow \alpha + A_\alpha \Delta\alpha, & \qquad & A_\alpha \in \{-1,0,1\} \\
\psi \rightarrow \psi + A_\psi \Delta\psi, & \qquad & A_\psi\in\{-1,0,1\} 
\end{array}
\end{equation}
The total number of possible actions $A=(A_\alpha,A_\psi)$ is 9 and the values of the increments $\Delta\alpha$ and $\Delta\psi$ are chosen based on the aerodynamic characteristic of the kite (see Supplementary information).

\subsection{The choice of observables}
The actual state $X$ of an AWE device immersed in a turbulent flow is an extremely high-dimensional vector as it includes all the degrees of freedom of the kite and the vehicle, as well as the wind field at any point in space and time. A direct approach is clearly unfeasible. It is therefore necessary to identify some relevant features, that is to map the Markov state $X$ into a smaller set of observables $S$, which can summarize the state of the system without losing valuable information. 

The selection of such observables could be in principle performed automatically. However this procedure requires very large training datasets that are very expensive to obtain. In addition, the selected features so obtained might be difficult to interpret in physical terms.  Here we take the alternate approach of choosing observations based on some physical intuition about which variables matter the most in achieving good control of the system. This approach requires expert domain knowledge and may fail if the observables are poorly chosen and convey little or no information about the state of the system. However, in our experience so far, the benefits of a physics-informed approach in terms of speed of learning, data parsimony and interpretability largely outweigh the limits in performance. 

Following this idea, we expect that the aerodynamic forces are a determinant factor for the ability of controlling the kite. As discussed in the previous section, lift and drag depend on the angle of attack $\alpha$, on the bank angle $\psi$, and on the relative velocity ${\bm v}_{rel}$.
It is therefore natural to consider them as relevant observables. In a further effort to compress the input we will consider just the orientation of the relative velocity with respect to the ground, given by the angle $\beta=\arcsin({v_{rel}^z}/\abs{{\bm v}_{rel}})$.
These variables are then discretized to obtain a finite set of observables 
\begin{equation}
   \begin{array}{l}
S_\alpha \in \{\alpha_{min}+ i\Delta \alpha\;; i=0,\ldots,N_\alpha \}\\
S_\psi \in \{\psi_{min}+ j\Delta \psi\;; j=0,\ldots,N_\psi \}  \\
S_\beta \in \{\beta_{min}+ k\Delta \beta\;; k=0,\ldots,N_\beta \}
 \end{array}
\end{equation} 
with a total number of observables $S=(S_\alpha,S_\psi,S_\beta)$ equal to $(N_\alpha+1)(N_\psi+1)(N_\beta+1)$ which in our learning experiments amounts to a few hundreds (see SI). 

In spite of the drastic reduction of dimensionality of the state space that is operated by this choice of observations, we will see that this information is sufficient to learn a very effective control strategy.

\subsection{The reward structure}
The last key ingredient is the reward, which is the numerical signal issued by the environment that provides feedback about the consequences of the performed actions. As the goal to be pursued is to maximise the traction power that we can extract from the wind, we use as a reward a close proxy that is the distance travelled along the mean wind in a time step \(\left|\dot{x}_v\right|\Delta t\).  In the Supplementary Information we explore different reward structures not necessarily aligned with the main component of the wind. When the kite crashes to the ground it receives a penalty, i.e. a negative reward, whose magnitude may depend on the time elapsed from the start (for instance, early falls are penalized more than late ones).
In addition, to avoid situations in which the kite flies dangerously close to the ground, we reduce the reward obtained whenever \( z_k \) goes below a threshold height \( z_{low} \). The detailed parameters used in the simulations are given in the Supplementary Information. 

\begin{figure*}[ht]
    \centering
    \includegraphics[scale = 0.72]{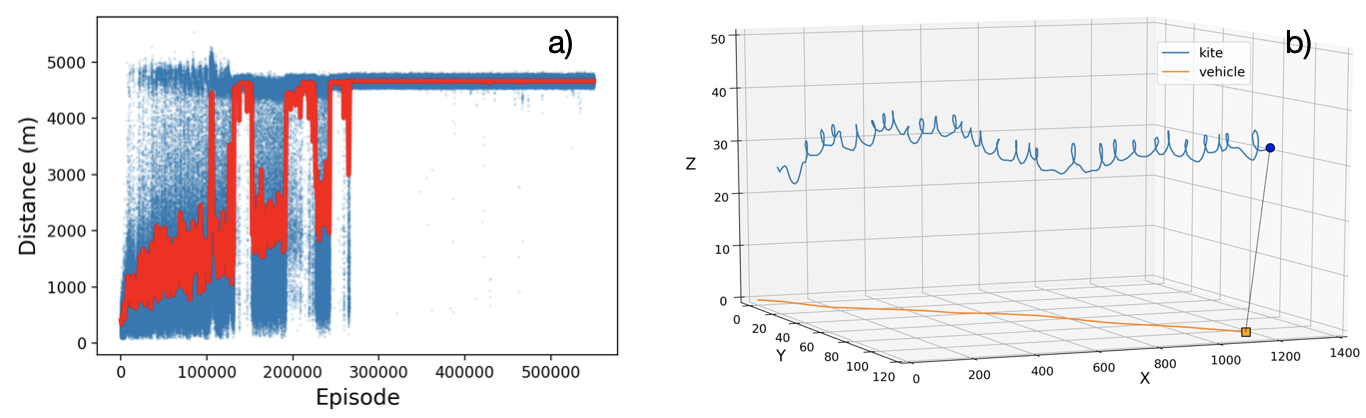}
    \caption{Discovering effective control strategies with Reinforcement Learning.  a) Horizontal distance covered by the vehicle as learning progresses. The blue dots refer to single episodes while the red line is a moving average over \(500\) episodes. Figure b) represents a sample of  learned motion in the turbulent Couette channel. The kite displays a helical motion adapted to the fluctuations of the wind flow. Note that the vehicle moves also in the \(y\) direction even if it is not directly rewarding since only the distance covered along \(x\) is accounted for in the return.}
    \label{fig:return}
\end{figure*}
\begin{figure*}[ht]
    \centering
    \includegraphics[scale=0.75]{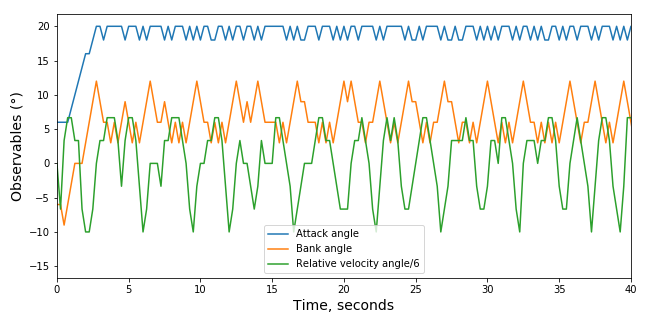}
    \caption{Dynamical behavior of the observables under the learned policy. Two stages emerge: a very brief transient where the kite reacts to the initial conditions, and an approximately periodic pattern where the attack angle remains close to its maximum values whereas the bank angle and the relative wind velocity angle oscillate out of phase.
     }
    \label{fig:pol_eval}
\end{figure*}

\section{Results}
In this section we present the results obtained by implementing the Reinforcement Learning algorithm that described above with special emphasis on the interpretation of the learned strategy in terms of simple decision rules.

\subsection{Learning effective control strategies}
The training is divided into episodes that terminate either when the kite crashes or after a sufficiently long time. 
At the beginning of each episode the vehicle is randomly initialized in a different point on the ground and the kite is at a given relative position from it. Random control angles are chosen in order to sample different flow conditions and kite postures, in order to obtain robust strategies and avoid overfitting. The system is motionless at time \(t=0\) for every episode, with fixed initial angles of the tether \(\theta\) and \(\phi\). Different take-off configurations can be considered depending on the specifics of the system at hand.

The initial estimates of the return \(\hat{Q}(s,a)\) for each state-action pair are chosen optimistically in order to favor exploration of the state-action space \cite{sutton}. 

The best results were obtained by scheduling the learning rate depending on the number of visits of the current state-action pair. This ensures faster updates of state-action values that have been visited less and vice versa (see Supplementary Information for details).

From Fig. \ref{fig:return}a we can see that after a few hundreds of thousands of episodes the distance covered along the \(x\)-axis converges to a stable value and maintains this performance for the remainder of the episodes. We have then evaluated the learned policy on a sequence of test episodes that are different from the training ones and, importantly, last longer. The fact that the performance is unaltered confirms that the learned strategy is able to generalize to previously unseen wind configurations (see Supplementary Information).

As shown in Fig. \ref{fig:return}b the trajectory of the kite has an approximately helical shape which changes over time depending on the local wind speed and direction. The towed vehicle moves along an approximately straight path with a sideways component with respect to the mean wind. We now turn our attention to the learned control strategy, uncover its main properties and distil a simple control strategy that achieves comparable performance.

\begin{figure*}[ht]
    \centering
    \includegraphics[scale=0.62]{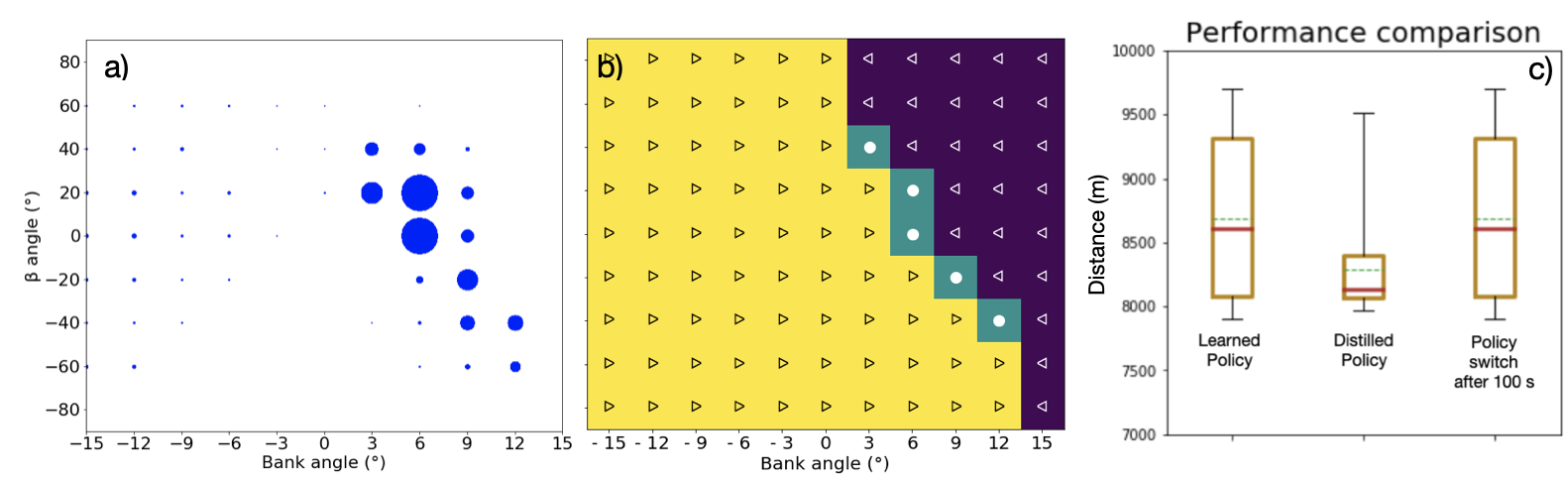}
    \caption{A distilled control strategy. a) Number of visits in the \(\beta_{t-1},\psi_t\) space highlighting the anticorrelation between the relative wind velocity direction and the bank angle. b) The distilled policy keeps the attack angle fixed and changes the bank angle as shown. Right arrows, left arrows and circles correspond to increase, decrease or maintain the bank angle, respectively. c) The performance of the learned policy and of the distilled one are compared, together with a mixed strategy that switches from the first to the second after \(100\,s\). Medians are in red, averages in dotted blue, the boxes are the quartiles. }
    \label{fig:3dtraj}
\end{figure*}
\subsection{Distilling a simplified control strategy}

Inspecting the evolution of the system under the learned policy, it is possible to identify two stages (Fig. \ref{fig:pol_eval}): a short transient that lasts from a few to some tens of seconds where the kite reacts to the initial motionless condition of the system, followed by a stable and approximately cyclic behavior in which bank angle \(\psi\) and \(\beta\) angles oscillate out of phase whereas the attack angle is nearly constant and close to its maximum value.

This observation suggests that the basic control mechanism underlying the helical motion of the kite can be actually explained in terms of simple control rules. A straightforward correlation analysis between \(\beta_{t-1}\) and \(\psi_t\), gives a Pearson correlation coefficient \(\simeq-0.82\). 
The time delay between \(\beta_{t-1}\) and \(\psi_t\) hints at the fact that the control reacts to changes in the direction of relative wind velocity by adapting the bank angle accordingly. The anticorrelation is also conspicuous when looking at the occurrence of visits in the 
\(\beta_{t-1}\) and \(\psi_t\) space, shown in Fig. \ref{fig:3dtraj}a. Building up on these considerations we defined a \textit{distilled policy} that keeps the attack angle fixed to the highest admissible value and changes the bank angle in such a way to reproduce the same pattern of visits as the one of the learned policy. This distilled control strategy is displayed in Fig. \ref{fig:3dtraj}b. 
This simple policy turns out to guarantee a performance comparable to the learned one (Fig. \ref{fig:3dtraj}c), The small reduction in distance traveled can be explained by the different ways in which the learned and the distilled policy manage the initial transient phase. Indeed, implementing an improved version in which the agent switches from the learned policy to the distilled one after some time achieves a performance that is indistinguishable from the fully learned one.



\section{Discussion}
We have shown how Reinforcement Learning can discover effective control strategies to manoeuver a kite to the end of providing traction power. The kite-vehicle system is immersed in a simulated environment that comprises the essential and unavoidable effect of atmospheric turbulence.

Our results have been obtained by an algorithm which only takes into account the control angles and the orientation of the relative wind velocity as observables. The learned control can be interpreted in terms of simple rules: 
\begin{itemize}
    \item[\it{i)}] keep the attack angle constant and as large as possible,
    \item[\it{ii)}] given a certain measurement of the relative wind velocity angle \(\beta\), increase or decrease the bank angle \(\psi\) in order to reach a target value that depends on \(\beta\) in an approximately linear way, with a negative proportionality coefficient (see Fig. \ref{fig:3dtraj}b).
\end{itemize}
How this strategy generalizes to other simulated environments with different velocity statistics remains an open question that we want to address in the near future. 

In our approach we selected the relevant observables based on our physical intuition. Another possibility would be to delegate the choice of the most appropriate features to the algorithm itself, for instance approximating the return  \(\hat{Q}\) by means of an artificial neural network as in Deep Q-Learning \cite{DRLreview}. The latter approach could in principle discover more appropriate inputs and lead to even better performance. However, this class of algorithms are infamously known to be very data thirsty. In addition, their results are often hard to interpret in terms of human-readable rules. Here we deliberately resolved the trade off in favor of rapid training and increased explainability rather than performance. It would nonetheless be of great interest to explore alternate approaches.

The application of our method beyond the simulated environment is a tantalizing perspective. However, several challenges lie ahead when training takes place in the real physical world. Among those, a prominent necessity is finding algorithms that learn faster. Encouraging results from robotics and unmanned aerial navigation, e.g. \cite{reddy2018glider}, offer some hope that these challenges can be overcome and that Reinforcement Learning can become an important algorithmic tool for AWE applications.

This project has received funding from the European Union’s Horizon 2020 research and innovation programme under the Marie Skłodowska-Curie grant agreement N°956457.

We are grateful to Lorenzo Basile and Emanuele Panizon for many profitable discussions and useful comments.

Nicole Orzan and Claudio Leone contributed equally to this work.

\bibliography{biblio} 
\end{document}